# A triple-mode mid-infrared modulator for all-surface radiative thermal management

Haoming Fang[1,2], Wanrong Xie[1,3], Xiuqiang Li[1], Kebin Fan[4], Yi-Ting Lai[1], Bowen Sun[1], Shulin Bai[2], Willie J. Padilla[4], Po-Chun Hsu[1,*]

[1] Department of Mechanical Engineering and Materials Science, Duke University, Durham, North Carolina 27708, USA

[2] Department of Materials Science and Engineering, HEDPS/CAPT/LTCS, Key Laboratory of Polymer Chemistry and Physics of Ministry of Education, College of Engineering, Peking University, Beijing 100871, China

[3] College of Materials Science and Engineering, Beijing University of Chemical Technology, Beijing 100029, China

[4] Department of Electrical and Computer Engineering, Duke University, Durham, North Carolina 27708, USA

Corresponding author: Po-Chun Hsu (pochun.hsu@duke.edu)

## SUMMARY

Thermal management is ubiquitous in the modern world and indispensable for a sustainable future. Radiative heat management provides unique advantages because the heat transfer can be controlled by the surface. However, different surface emissivities require different tuning strategies. Here, we demonstrate a triple-mode mid-infrared modulator that can switch between passive heating and cooling suitable for all types of object surface emissivities. The device is composed of a surface-textured infrared-semi-transparent elastomer coated with a metallic back reflector, which is biaxially strained to sequentially achieve three fundamental modes: emission, reflection, and transmission. By analyzing and optimizing the coupling between optical and mechanical properties, we achieve a performance of emittance contrast $\Delta\varepsilon$ = 0.58, transmittance contrast $\Delta\tau$ = 0.49, and reflectance contrast $\Delta\rho$ = 0.39. The device can provide a new design paradigm of radiation heat regulation to develop the next generation wearable devices, robotics, and camouflage technology.

# INTRODUCTION

Thermal management is widely used and critical in advanced technology[1–3]. For instance, batteries, solar cells, vehicles and electronic chips all require careful thermal management to maintain suitable working temperature ranges[4–10]. On a larger scale, space heating and cooling maintain the indoor temperature all year round but also consume up to 20% of global energy[11,12]. Therefore, the development of more effective, energy-efficient, and universal thermal management technology is one of the most critical tasks for a sustainable future[13–15].

Among all the thermal management mechanisms, radiation heat transfer is best known for its universality, media-free operation, and large tunability[16,17]. Materials with exotic radiative heat transfer properties were developed through photonic design and fabrication[18–20], and they have shown great promise as an energy-efficient, or even an energy-free approach, to reduce the thermal management energy demand[21,22]. Fundamentally, the radiative property of any material has three components: reflectance ($\rho$, including both specular and diffuse), transmittance ($\tau$), and absorptance ($\alpha$), which sum to unity, i.e., $\rho + \tau + \alpha = 1$. According to Kirchhoff's Law, at thermodynamic equilibrium, the absorptance equals the emittance ($\varepsilon$), $\alpha = \varepsilon$. Therefore, for radiatively opaque ($\tau = 0$) materials, $\alpha = \varepsilon = 1 - \rho$, and the total power ($P$) radiated from an object is given by Stefan-Boltzmann Law: $P = A\varepsilon\sigma T^4$, where $A$ is surface area, $\sigma$ is Stefan–Boltzmann constant and $T$ is the object surface temperature. As shown in Figure 1A, to achieve radiative cooling for low-ε objects such as metals, the strategy is to apply a high-ε coating to promote thermal emission. For high-ε objects, such as the human body, the ideal cooling strategy is to cover them with high-$\tau$ materials allowing direct radiation from the skin surface. On the other hand, the radiative heating strategy of applying low-ε opaque materials works for both low-ε and high-ε objects[23]. The dependence of cooling strategies on the object emissivity is further elaborated in Figure 1B and 1C. When a high-$\tau$ material is used for cooling, the heat transfer coefficient increases approximately linearly with the underlying object emissivity. For a high-ε material, the heat transfer coefficient depends only on the cooling material itself, but its thermal resistance prohibits it from achieving better cooling when the object emissivity is already high. Besides, the size of the air gap between object and film also plays an important role in thermal transfer performance. (The detailed analysis of these hypothetical cooling materials and thermal circuit are shown in the Supplementary Text.) Taking the airgap of 1 mm as an example (Figure 1D), one can determine the high-$\varepsilon$-dominating and the high-$\tau$-dominating regions for various object emissivities. This result unambiguously shows the demand for the dynamic regulation among all three modes—emission, reflection, and transmission—is necessary to achieve effective radiative thermal management for an arbitrary object[24].

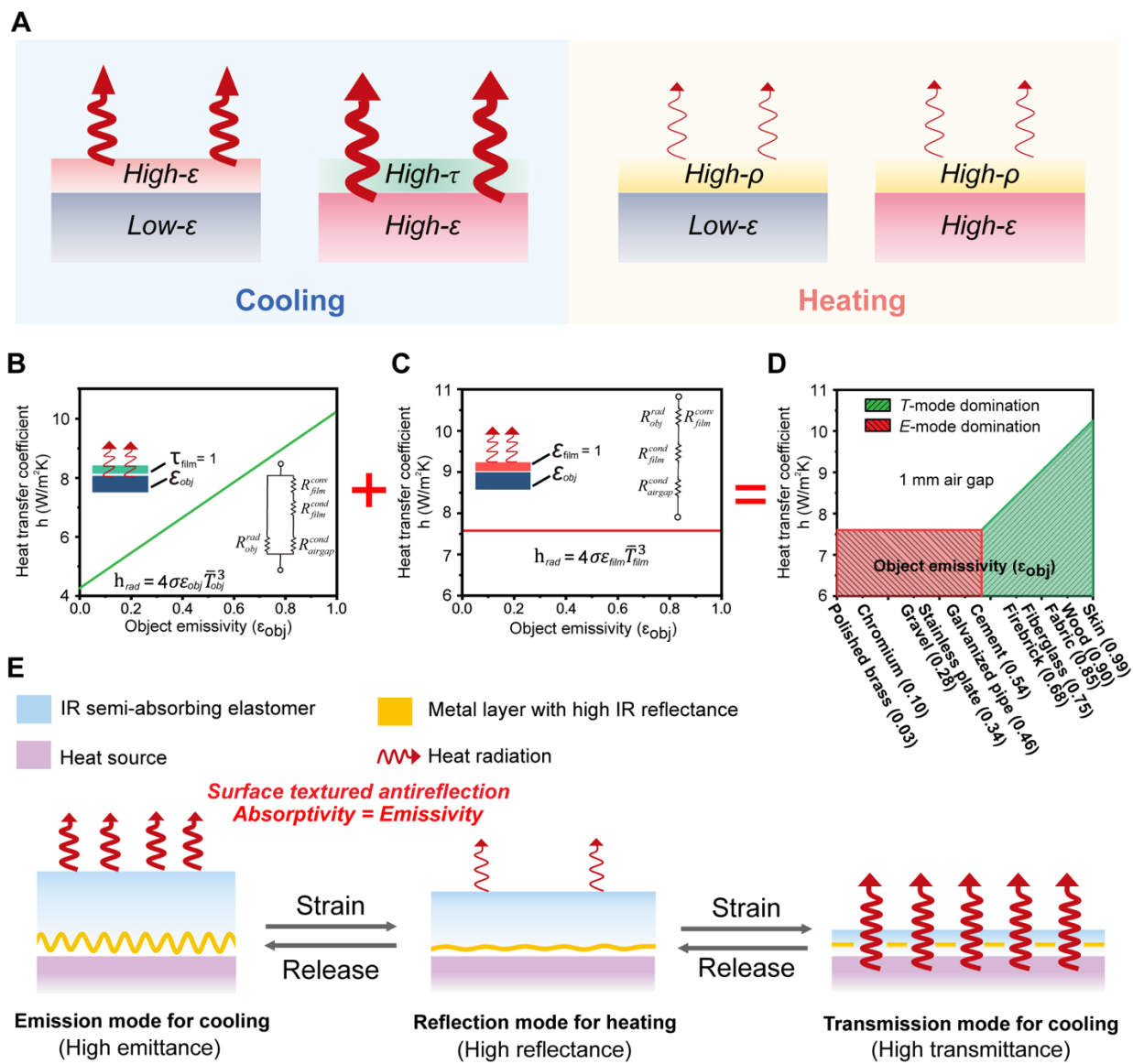

**Figure 1. Design principle of the triple-mode mid-IR modulator for all-surface thermal management.**

**(A)** For cooling, the best strategies for high-emissive (high-$\varepsilon$) and low-emissive (low-$\varepsilon$) objects are by materials with high-transmittance (high-$\tau$) and high-emittance (high-$\varepsilon$), respectively. For heating, using high-reflective (high-$\rho$) material works for all objects.
**(B)** When the radiative cooling strategy is by IR-transmittance, the radiative heat transfer coefficient is dependent on the object's emissivity, which is less effective for low-emissivity objects.
**(C)** When the radiative cooling strategy is by IR-emittance, the radiative heat transfer coefficient is constant but underperforms for high-emissivity objects.
**(D)** To achieve optimal radiative cooling on all kinds of object emissivities, the thermal-managing material must be able to switch between high transmittance and high emittance.
**(E)** Based on optical and mechanical design, the thin film mid-IR modulator can change among "*Emission-mode*", "*Reflection*-mode", and "*Transmission*-mode" via mechanical actuation.

Here, we demonstrate a triple-mode polymer/metal elastomeric modulator that is able to dynamically switch among emission (*E-mode*), reflection (*R*-mode), and transmission (*T*-mode) to accomplish the management of the all-surface passive radiative cooling and heating. As shown in Figure 1E, in *E*-mode, the IR semi-absorbing elastomer is at zero strain and the largest thickness for absorption. The underlying metal layer has the highest roughness to promote diffuse scattering and further increases the absorption[25]. In *R*-mode, the film is biaxially stretched, which smoothens the metal reflector and shrinks the elastomer thickness for high reflection. In *T*-mode, the film is stretched further to reduce the thickness and create voids on the metal film that results in high transmittance.

## Results

**Triple-mode modulator design principle.**

The realization of the triple-mode modulator calls for a rational design of both the mechanical properties and the optical structure[26–31]. For the elastomeric superstrate, the styrene–ethylene–butylene–styrene (SEBS) block copolymer is chosen because of its excellent mechanical properties, thickness-sensitive mid-IR absorption, and solution-processability. In contrast, polyethylene (PE) is IR-transparent but non-elastic, and polydimethylsiloxane (PDMS) elastomer is IR-opaque (Figure S2). Figure 2A illustrates the fabrication procedure. First, a 5 wt% SEBS/hexane solution is drop-casted onto a surface-textured silicon wafer to create a highly diffuse surface – scanning electron microscope (SEM) images are shown in Figure S4. The SEBS film was then delaminated from the wafer and biaxially stretched to 100% strain, followed by a deposition of 70-nm-thick Ti/Au as the IR-reflective layer. At this tensile strain state, the film works in *R*-mode because the metal layer is at its lowest roughness state. After releasing the strain, the SEBS thickness increases and the metal layer return to the highly-textured state, and the film is in *E*-mode. When the film is stretched to 200% strain, the metal layer becomes island-like and operates in *T*-mode. The three modes are characterized in the visible, an SEM (from the metal side), and by an optical profiler (from the SEBS /metal interface), as shown in Figure 2B, C, and D, respectively. In *E*-mode, the film is dark due to the anti-reflecting surface texture that has an RMS roughness of 1.632 µm. When it is adjusted to *R*-mode, it becomes visually reflective, and the roughness decreases to 0.997 µm, showing the metal layer is smoothened. Further stretching the film to *T*-mode, at which it becomes visually transparent because the metal domains are separated from each other.

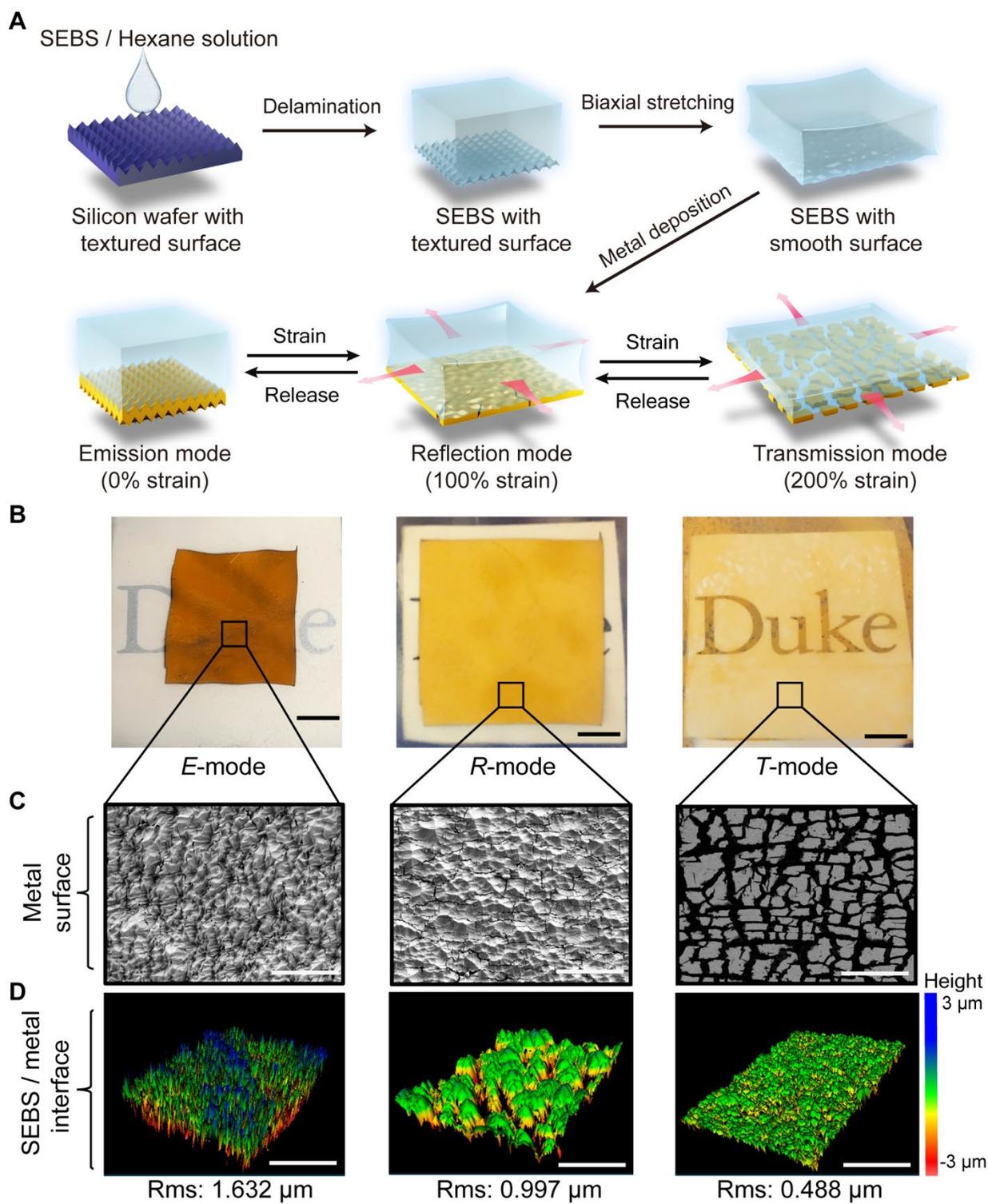

**Figure 2. Fabrication and morphology characterization of the triple-mode modulator.**
**(A)** Schematic of film preparation method and the mechanical actuation of the triple-mode modulator with biaxial strains of 0% (left), 100% (middle) and 200% (right).

**(B)** Digital camera images of the film under the biaxial strains of 0% (left), 100% (middle) and 200% (right). The scale bars are 5 mm.
**(C)** Scanning electron microscopy (SEM) images of a representative sample from the metal side under the biaxial strains of 0% (left), 100% (middle) and 200% (right). The scale bars are 50 μm.
**(D)** The 3D topographical maps of the sample surface from the SEBS side under the biaxial strains of 0% (left), 100% (middle) and 200% (right). The scale bars are 50μm.

**Strain-emittance coupled optimization.**

The emittance of SEBS/metal film is determined by both the SEBS thickness and the metal diffuse scattering[32,33], as shown in Figure 3A. A detailed analysis of these two factors is performed for maximizing the emittance contrast in the modulation range. We define the total effective absorption coefficient ($\mu$) to be the product of the intrinsic absorption coefficient ($\mu_p$), and the enhancement factor ($\beta$) which is attributed to the increased optical path length compared with the planar back reflector. The transmittance of a smooth SEBS film without the metal back reflector is described by the Beer-Lambert law as $\tau = (1-R)^2 \exp(-\mu l) = (1-R)^2 \exp(-\mu_p d)$ where $l$ is the light path length, $d$ is the nominal film thickness, and $R$ is the Fresnel reflection at the interface. By measuring the transmittance of smooth SEBS films without the metal back reflector (Figure S5), we determined $\mu_p$ = 0.00377 μm$^{-1}$ (weighted-averaged over room temperature black body radiation spectra from 8 to 12 μm). Next, the $\beta$ values are determined by measuring the reflectance $\rho$ of SEBS/Au thin films of different strains, thicknesses, and surface morphologies. As shown in Figure 3A, the black curve represents the sample without surface texturing ($\beta_{0\%} = \beta_{100\%} = 1$), and the maximum emittance difference ($\Delta\varepsilon_{max}$) is 0.36 when the film thickness is approximately 200 μm at zero strain. When the inverse-pyramid texture is introduced to the SEBS film, $\Delta\varepsilon_{max}$ increases to 0.43 at a thickness of 125 μm, which corresponds to the enhancement factor $\beta_{0\%}$ = 1.9 at 0% strain and $\beta_{100\%}$ = 1.5 at 100% biaxial tensile strain. Although other thicknesses of SEBS would also have the emissivity-reflectivity modulation, our semi-empirical analysis provides the guideline for the materials design which directly results in superior tunability. It is also worth noting that biaxial stretching actuation has better performance than the simple uniaxial and side-constrained uniaxial stretching methods (More detailed discussion was shown in Supplementary Text). Apparently, $\Delta\varepsilon$ will further increase with higher ($\beta_{0\%} - \beta_{100\%}$) in Figure 3C, assuming $\beta_{0\%} = 1$, as $\beta_{100\%}$ increases, $\Delta\varepsilon_{max}$ increases, and its corresponding thickness decreases. The theoretical upper limit of $\beta_{0\%}$ (and therefore the limit of $\beta_{0\%} - \beta_{100\%}$ and $\Delta\varepsilon$) occurs for a perfect Lambertian surface which is known as the Yablonovitch limit. Based on our definition of $\beta$, this upper limit dictates $\beta \leq 2n^2$, where $n$ is the reflective index of the IR absorber. Inserting $n_{SEBS} = 1.56$ yields $\Delta\varepsilon_{max}$ to be as high as 0.72 (Figure 3C) when $\beta_{0\%} = 4.7$ and $\beta_{100\%} = 1$, suggesting the promising modulation range between E- and R- mode via better optical structure design and engineering[34,35].

It should be noted that the elastic strain-induced light modulation has been extensively studied with promising functionalities and performances[36]. To the best of our knowledge, most prior reports focused on the wavelength regions of visible and near-IR (except for reference[27,30,37]), which have drastically different material property requirements from mid-IR region. While we received inspiration from these literatures, our mid-IR triple modulator is the outcome of several new critical designs. First, we choose SEBS over PDMS because SEBS has the proper mid-IR absorption coefficient, and PDMS is too absorbing to be mechanically tunable with a proper thickness. Second, the additional surface texturing is the critical and dominating factor for large emissivity contrast between E- and R-mode, which is in contrast with the interfacial wrinkling due to moduli mismatch. In Figure 3B, the SEBS/Au sample without surface texturing (marked as green squares) has an emissivity contrast of only 0.31, which agrees well with the smooth surface scenario. Surface texturing increases the Δε to 0.41. This implies that, without the additional surface texturing, the strain-induced wrinkles do not have enough roughness to produce mid-IR diffuse scattering, and almost all the emissivity contrast (Δε = 0.31) is caused by the thickness change. These two designs are the key factors for high modulation performance and are distinct from the visible/near-IR modulation.

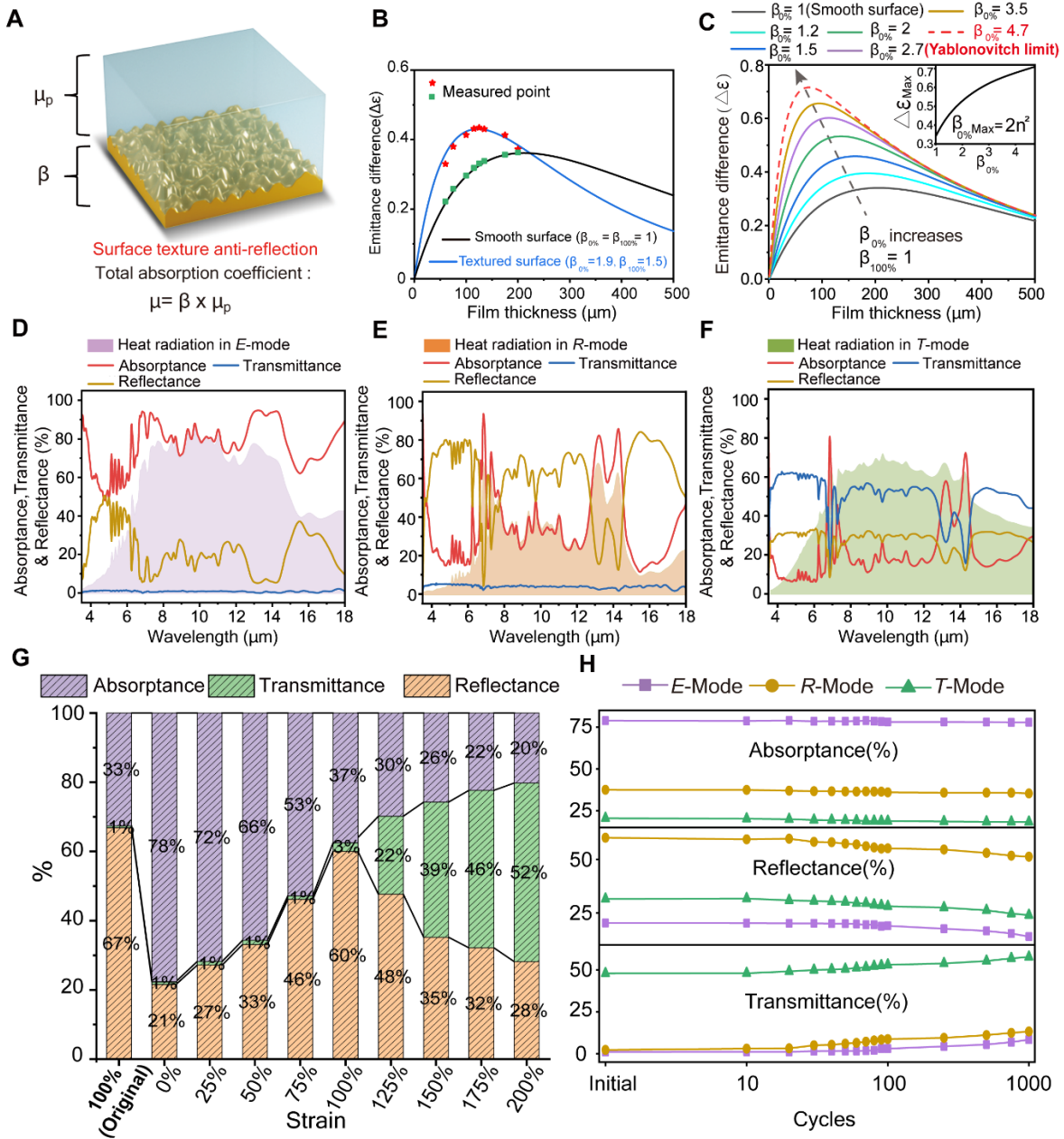

**Figure 3. Infrared properties and strain-emittance coupled optimization.**

**(A)** Schematic of the sample model which consists of total absorption coefficient $\mu$, intrinsic absorption coefficient $\mu_p$ and enhancement factor $\beta$.

**(B)** The emittance difference between *E*- and *R*-mode in smooth (black line) and textured (blue line) SEBS film as a function of film thickness. The red stars and green squares represent the measured values.

**(C)** The emittance difference with increasing $\beta_{0\%}$ and fixed $\beta_{100\%} = 1$ as the function of film thickness. The inserted plot is the maximum emittance difference at different $\beta_{0\%}$.

**(D,E,F)** The absorptance, transmittance, and reflectance of the triple-mode modulator with the strain of 0% (*E*-mode), 100% (*R*-mode), and 200% (*T*-mode) in the infrared spectra from 3.5 to 18 μm. The shaded areas represent the resultant heat radiation.
9

**(G)** Weighted-average absorptance, transmittance, and reflectance of the film under the biaxial strain from original 100% to 0% and then stepwise to 200%.
**(H)** 1000 times cycle stability test of the weighted-average absorptance, transmittance, and reflectance of E-, R-, and T-mode.

**Infrared modulation**.

In *E*-mode (Figure 3D), the modulator has the highest absorptance (emittance) of 78 ± 1% and low transmittance of ~1% owing to the corrugated metal as well as the largest SEBS thickness. In *R*-mode (Figure 3E), the reflectance dominates with a value of 60 ± 1%, and the transmittance becomes ~3 ± 1%. The slight increase in transmittance is caused by the voids between the metal plates. In *T*-mode (Figure 3F), the transmittance of the film rises to 52 ± 2%, while the reflectance and absorptance decrease to 28 ± 2% and 20 ± 1%, due to the isolated metal islands and the reduced SEBS thickness. The strong absorption peaks at the wavelengths of 6.3 μm, 6.8 μm, 7.2 μm, 9.7 μm, and 13.3 μm are attributed to the functional groups of SEBS such as –C=C–, –CH$_2$–, and =C–H[38]. The shaded regions in each mode are the approximate heat radiation, suggesting the correlation between mid-IR modulation and radiative heat management. Figure 3G shows these three mid-IR properties after weighted-averaging over black body radiation when the strain varies from 0% to 200%. From *E*- to *R*- mode, the reflectance increases by 39% with the same amount of emittance decrease. From *R*- to *T*- mode, the transmittance dominates the thermoregulatory performance by a 49% increase, and both reflectance and emittance decrease. This exceptional tunability is stable after 1000 strain cycles with around 10% change, indicating the not bad cycle stability and reversibility when comparing with the reference in Table S9 (More detailed discussion was shown in Supplementary Text). (Figure 3H and the stability failure analysis is revealed via SEM images from Figure S23 to Figure S25.) The corresponding reflectance spectra including specular and diffuse reflectance among three modes are shown in Figure S19.

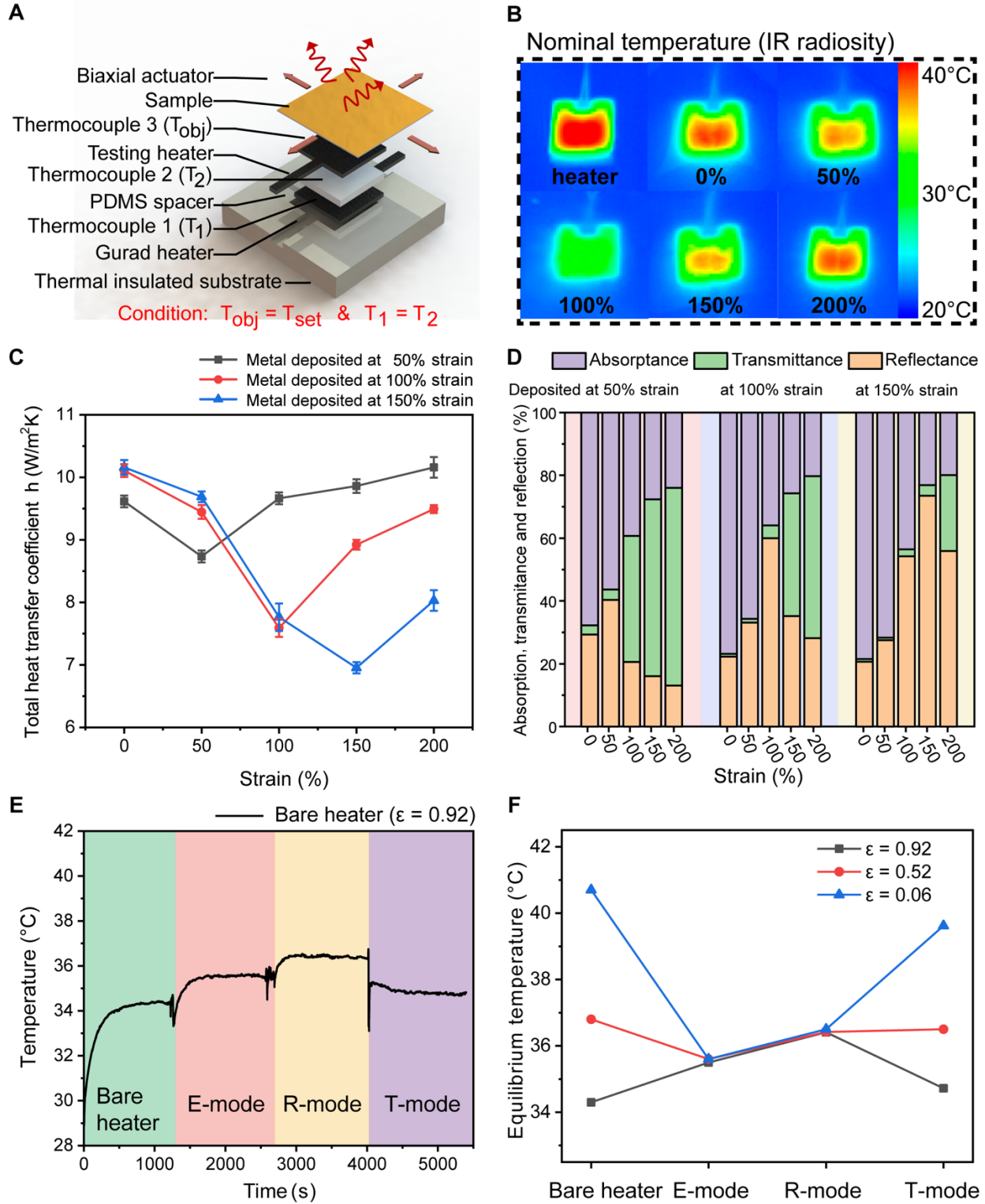

**Figure 4. Thermal measurement of the tri-mode film.**
**(A)** Side-view of steady-state measurement apparatus.
**(B)** The top surface infrared images of testing heater and sample under 0%, 50%, 100%, 150%, and 200% biaxial strain.

**(C)** Total heat transfer coefficient of the film with metal deposited under 50% (dark grey line), 100% (red line) and 150% (blue line) biaxial strain as a function of strain.
**(D)** The corresponding weighted-average absorptance, transmittance, and reflectance of the films with metal deposited under 50%, 100% and 150% biaxial strain.
**(E)** The actual temperature change among E-, R- and T-mode. (The SEBS/metal film with metal deposited under 50% biaxial strain)
**(F)** Summary of equilibrium temperature at different states. (Black line is the heater with an emissivity of 0.92, red line is the heater with an emissivity of 0.52, and blue line is the heater with an emissivity of 0.06)

**Thermal measurement**. The radiative heat transfer management of the triple-mode modulator was validated via a steady-state measurement apparatus as shown in Figure 4A and Figure S28. As shown in Figure 4C (red line), the total heat transfer coefficients of the triple-mode modulator are measured to be 10.11 ± 0.11 W/m$^2$K, 7.59 ± 0.14 W/m$^2$K and 9.49 ± 0.07 W/m$^2$K for *E*-, *R*-, and *T*-mode, respectively. The heat transfer modulation range and the trend can be further controlled by depositing the metal back reflector at different starting strains. Again in Figure 4C, heat transfer coefficients of modulator whose metal deposited at 50% and 150% strains were investigated and shown in dark grey and blue line, respectively. The weighted-averaged mid-IR properties of these three samples are also measured and shown in Figure 4D This demonstrates the additional degree of freedom to design and customize the triple-mode radiative heat management device to suit various kinds of scenarios and environment requirements[39,40]. The infrared images (Figure 4B) show the nominal temperature (IR radiosity) distributions of the film under different strains. Further, the actual temperature change of object/heater among E-, R- and T-mode presents the large thermoregulation with up to 6 degrees (Figure 4E, F). Notably, if our device is used for indoor personal thermal management, such temperature difference can result in an energy saving of several tens of percentage[41]. It is worth mentioning that if the emissivity of the heater is higher than our modulator, the lowest temperature would be obtained in the bare heater as shown in Figure 4E. However, for the low-e heater with an emissivity of 0.06, there is an obvious temperature drop of 6 degrees when applying our modulator in E-mode, which demonstrates the cooling effect. Again, the necessity of three modes for various objects with all emissivity was demonstrated in Figure 4F and Figure S30.

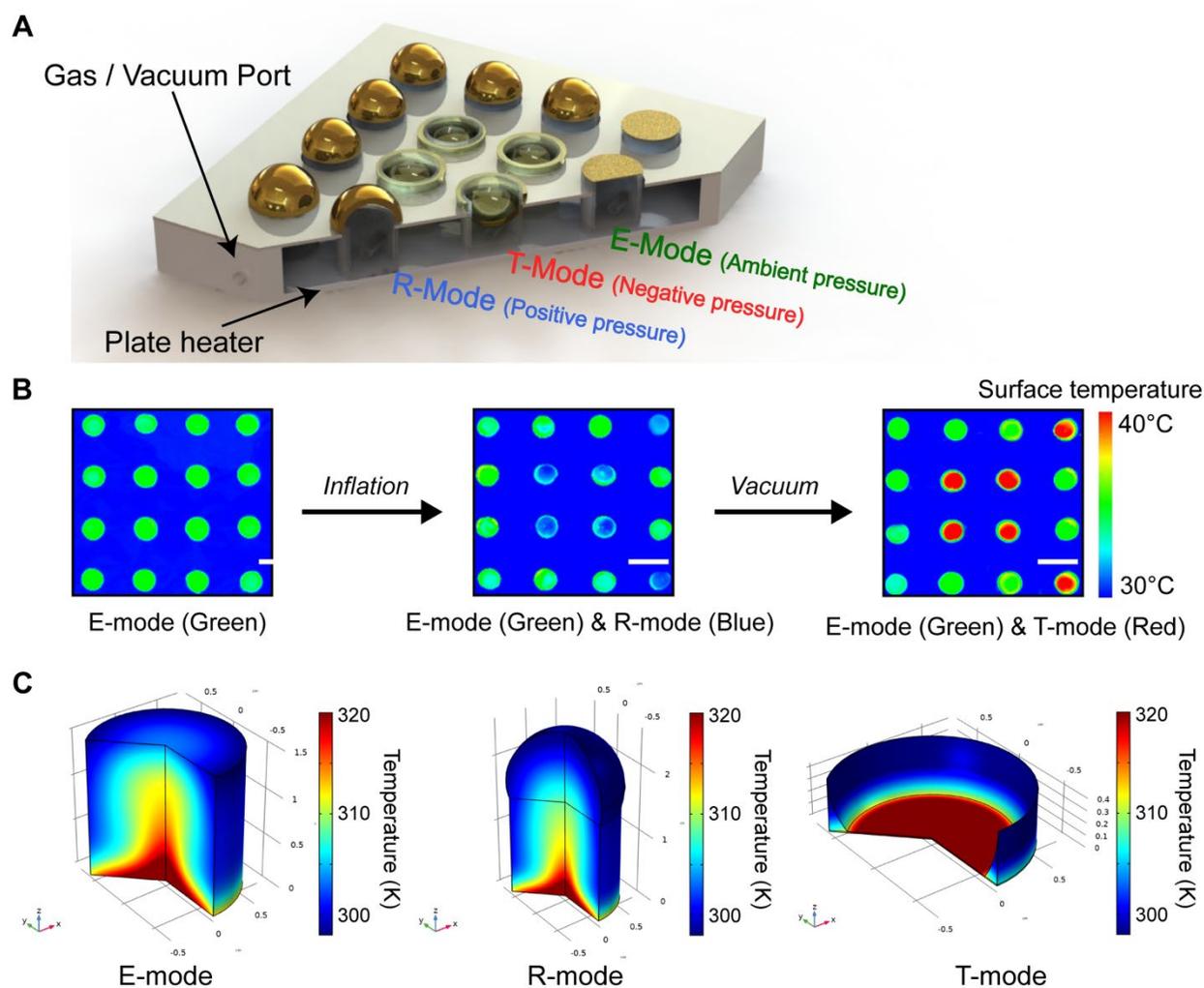

**Figure 5. Pneumatic modulation of the infrared appearance for 4 x 4 pixels device.**

**(A)** Schematic of the pneumatic device with a 4 x 4 pixels. *E*-mode is at ambient pressure, *R*-mode is at positive pressure and *T*-mode is at negative pressure.

**(B)** Infrared image from top-view. The green dots are in *E*-mode, the blue dots are in *R*-mode and the red dots are in *T*-mode. The scale bars are 2 cm.

**(C)** Temperature distribution of pixels in *E*-mode (left), *R*-mode (middle) and *T*-mode (right).

**Camouflage device by pneumatic control.**

In addition to heat management, a pneumatic thermal camouflage device with a 4 x 4 matrix is fabricated as shown in Figure 5A and Figure S32. By regulating the air pressure, the SEBS/metal film on each pixel can change among *E*-, *R*- and *T*-mode, further displaying different infrared appearances. The flat configuration is in *E*-mode. Under positive pressure, the film bulges into a dome-shaped cap and is stretched to *R*-mode. Under negative pressure, the film was depressed and stretched to *T*-mode, allowing the heat dissipation to the ambient. This pneumatic soft actuation[42] enables rapid and highly reversible thermal radiation control and combines both functions of infrared camouflage and heat management. Among *E*-, *R*- and *T*-mode, the apparent temperature contrast between pixels is approximately 5 °C. We further individually control the pressure of the pixels to show its capability as a thermal radiation "display". Initially, all pixels are in *E*-mode showing green color (Figure 5B, left). Inflating the six dots to *R*-mode changed the color would show a blue letter "D" (Figure 5B, middle). On the other hand, vacuuming the six pixels to *T*-mode would show a red letter "D" (Figure 5B, right). The relationship between air pressure and film deformation was calculated in supplemental text. To explain the camouflage devices mechanisms considering the radiative heat transfer and convection and conduction, we have used COMSOL multiphysics to simulate the temperature distribution inside the pixels (in Figure 5C) and calculated the radiosity among these three modes (in Table S7). We designed our pneumatic camouflage device to take advantage of both radiative heat transfer and convection/conduction. As shown in Figure S35B, there is a remarkable tunability in radiosity using the metal/SEBS modulator among these three modes. For R-mode, both the large airgap with low thermal conduction and the low surface emissivity induced by metal/SEBS modulator will result in low surface radiosity and disguise as a "cold object". For T-mode, the narrow air gap and high transmittance enable radiosity and pretend as a "hot object".

## DISCUSSION

In summary, the triple-mode mid-IR modulator has demonstrated its capability to accomplish the all-surface passive radiative cooling and heating management and mid-infrared camouflage. It is the first experimental demonstration of a triple-mode mid-IR radiative heat managing device, together with numerical modeling as theoretical support. Distinct from previous approaches that focus on static, single- or dual-mode heat management, our work provides several advantages for thermal management. First, as shown in Figure 1D and Figure S1C, when the goal is to provide cooling for objects with high emissivity (e.g. human body) and/or the device has appreciable air gap resistance, T-mode works the best. When the objects have low emissivity (e.g. metals) and/or the air gap is small, E-mode can provide better cooling. When heating is desired, R-mode always works. Such complete, all-scenario radiative heat management can only be obtained by the triple-mode device that can switch among T-, E, and R-doiminant modes. Second, the investigation of emissivity-strain correlation among tuning range, film thickness, and scattering roughness can serve as a useful guideline for the research community for further development. Third, the straightforward mechanically-driven actuation in our modulator provides real-time adaptation. The thermal response time is determined by the characteristic thermal diffusion time constant $\tau = \frac{L^2}{\alpha} \approx \frac{(100 \times 10^{-6} \text{m})^2}{10^{-6} \text{m}^2 \text{s}^{-1}} = 0.01 \text{ s}$, where we assumed the SEBS film thickness (L) is 100 μm and thermal diffusivity (α) is $10^{-6}$ $m^2s^{-1}$. This short time constant can provide quick adaptability to the needs. Fourth, although Au was used as the reflective metal layer, the selection of metal is not limited to Au. Other IR-reflecting metals can also work well with much lower cost and weight (Figure S9).

Besides mechanical actuation, we envision the triple-mode mid-IR modulation concept can also be accomplished by other methods such as thermochromic phase transition induced infrared modulation[43–45], electrochemical induced modulation[46,47] of the infrared reflectance, ultraviolet (UV) light-induced modulation[48], and liquid induced modulation[49,50], with different pros and cons and therefore suitable applications (Table S8). It should be noted that different applications have different additional engineering constraints to overcome. For example, for outdoor applications, the solar heat gain, the atmosphere transparent window, and convective loss can all affect the heating/cooling performance[14]. For human body thermoregulation, creating nanoporous structure in the film would be beneficial for moisture management. Detailed future investigation on more complicated thermal environment will be worthwhile. For IR camouflage, thermal metamaterials[51–55] with judiciously engineered thermal conductance show great promises by concealing the object without a priori knowledge of the background thermal radiation. It is anticipated that the incorporation of metasurface design and dynamic materials tuning can bring substantial synergistic benefits. With our proof-of-concept demonstration, many other exciting future works can be pursued with multidisciplinary efforts, such as plasmonic metamaterials, photonic crystals, device miniaturization, artificial intelligence-enabled camouflage, and integrated wearability. Overall, we envision this versatile and effective approach will endow immense opportunities for photonics, radiation heat transfer, robotics, and energy science.

# EXPERIMENTAL PROCEDURES

**Fabrication of SEBS/metal film.**

Firstly, a silicon wafer was selectively etched by 5 M potassium hydroxide/1 M isopropyl alcohol-water solution at 60 °C for 20 h and formed the pyramid structures as shown in Fig S4(a). Subsequently, the 5 wt% Polystyrene-block-poly (ethylene-ran-butylene) -block-polystyrene (SEBS) power (Sigma, the average molecular weight of 89,000) was dissolved in hexane and poured onto the glass molds with etched wafers as the substrate. To obtain the precise thickness, the height of the glass molds was well-controlled. Besides, the Petri dishes (diameter of 15 cm) serve as covers to lower the evaporation rate of hexane solvent and prevent the bubbles. After the SEBS film was dried up at room temperature for 24 h, it was delaminated from the wafer template and baked in a blast oven at 80 °C for 12 h to remove residual solvent. Last, the SEBS films were biaxial stretched to different strains (typically, 0%, 50%, 100%, 150%, 200%) via a lab-made actuator (Actuonix Linear Actuators, P16). The stretched SEBS films were oxidation treated via Plasma Asher under $O_2$ atmosphere for 30 s (Emitech, K-1050X) and deposited 10 nm titanium as an adhesive layer and 60 nm gold as a reflective layer by magnetron sputtering (Kurt Lesker PVD 75). The thicknesses of the metal layers were controlled by the sputtering power and time.

**Microstructure characterization.**

The scanning electron microscope (SEM) images were taken by FEI Apreo S (5 KV). Typically, the film was biaxial stretched to setpoint displacement via an actuator controlled by a LabVIEW program. Subsequently, the film was mounted by a specific clamp (Figure S29) under various strains and taken off from the actuator for imaging.

**Roughness measurement.**

The surface roughnesses of the film at different modes were characterized via 3D Optical Profiler (Zygo NewView 5000). Firstly, the SEBS film was stretched to the strain of 0%, 100% and 200% via a biaxial actuator and fixed with a clamp. A thin Au layer (around 15 nm) was sputtered onto the textured surface side of the film and served as the reflective layer for imaging. The 3D topographical images of the sample surface were obtained via white light interferometry. The microscope objective was set as 10X. The surface roughness and surface flatness were calculated by the topographical map.

**Thickness measurement.**

The thicknesses of the samples were measured by Bruker Dektak 150 profilometer. Typically, the film sample was placed onto a silicon wafer substrate, then stretched to a certain strain and fixed by four clips at each edge. A diamond-tipped stylus moves across the sample, surface variations cause it to be translated vertically. This corresponds to an electrical signal that gives feedback on the appropriate step height. Compared with the height of the substrate, the thickness of the sample was obtained.

**Mechanical test.**

The mechanical properties of the SEBS sample were characterized according to the standard test method (ASTM D638) via Lloyd Instruments Mark-10. The samples were dumbbell shape with 9.5 mm wide and 63.5 mm long and the gauge distance was 9.5 mm long. The displacement rate was 0.5 mm/min. The engineering stress ($\sigma$) was calculated as $\sigma = F/A$, where F is the applied force and A is the cross-section area of the sample. The engineering strain ($\varepsilon_e$) was calculated via $\varepsilon_e = \Delta L / L$, where $L$ is the actual length of the sample, $\Delta L$ is the displacement. The Poisson ratio was calculated by $\nu = L\varepsilon_z / L\varepsilon_x$, where $\varepsilon_z$ is the longitudinal strain and $\varepsilon_x$ is the transverse strain. The mechanical stability of the sample was tested under six loading cycles at 200% uniaxial strain with a displacement rate of 0.5 mm/min.

**Infrared properties.**

The reflectance ($\rho$) and transmittance ($\tau$) of samples from 3.5 μm to 18 μm were measured by an FTIR spectrometer (model 6700, Thermo Fisher Scientific) accompanied by a diffuse gold integrating sphere (PIKE Technologies). During the experiments, a SEBS/metal film was stretched to different strains and mounted with the clamp. The tested area of clamps is guaranteed to be large enough to totally cover the 2 cm port of the integrating sphere. The total reflectance spectra were obtained at an illumination angle of 12°, and the total transmittance spectra were collected at a normal illumination angle. The absorptance ($\alpha$)/emittance ($\varepsilon$) was calculated on the basis of $\alpha = \varepsilon = 1 - \rho - \tau$. The approximate heat radiations ($P$) in *E*-, *R*-, *T*-mode were calculated by $P = P_{bb} \times (1 - \rho)$ where $P_{bb}$ is the room temperature black body radiation spectra. The angular dependent reflectance spectra at 0% strain was characterized by an FTIR spectrometer (Bruker Vertex 80v). The sample was mounted vertically on a Harrick variable angle reflection stage with the incidence polarization controlled by a wire-grid polarizer. The angles were swept from 25° to 55° with a step of 10° for both TE and TM polarizations. After the sample characterization, reflections of a gold mirror with similar size were also measured at the same angles and same polarizations for references.

**Visible properties.**

The visible reflectance and transmittance of samples were measured by UV–visible–near-infrared spectroscopy spectrometer with a calibrated $BaSO_4$ integrating sphere (400–1600 nm, Agilent technologies, Cary 6000i). The absorptance was calculated on the basis of reflectance and transmittance.

**Cycling stability testing.**

The cycling stability of the sample was conducted via the biaxial actuator controlled by an automatic cycling LabView programming. Absorptance, reflectance, and transmittance of the sample were measured in the *E*-, *R*-, *T*-mode (strain = 0, 100, and 200%) every 10 cycles using the aforementioned method.

**Environmental stability testing.**

We use the IEC60068-2-67 standard as the environment stability testing principle. The film was put into the environmental test chamber (Espec, SH-222, Japan) and the temperature and humidity were set to be 85 °C and 85 RH. Absorptance, reflectance, and transmittance of the sample were measured in the E-, R-, T-mode (strain = 0, 100, and 200%) every 24 h using the aforementioned method.

**Thermal measurement.**

The total heat transfer coefficient was measured via a constant heat flux apparatus equipped with a testing module and a biaxial tensile module (Figure S28). The testing module consists of the test heater, spacer, guard heater, and thermocouples. The test heater was a flexible polyimide heater plate (1 inch x 1 inch, OMEGA Engineering, KHLVA-101/10) coated with a black acrylic layer ($\varepsilon \sim 1$). Two K-type thermocouples (SKA1, OMEGA Engineering) were attached to the center of the top and bottom surface in the testing heater, denoted as $T_{obj}$ and $T_2$. Same sized PDMS spacer and a guard heater (1 inch x 1 inch, OMEGA Engineering, KHLVA-101/10) were sequentially placed below the testing heater. Another K-type thermocouple (SA1-K, OMEGA Engineering) was attached to the center of the top surface of the guard heater, denoted as $T_1$. All these testing module components are sealed with thick latex foam to avoid heat leakage. We used an acrylic chamber covered by black aluminum foils ($\varepsilon \sim 1$) as the testing enclosure. A circulating water channel (Cole-Parmer) was fixed onto the chamber walls to regulate the ambient temperature. A K-type thermocouple (SA1-K, OMEGA Engineering) was located above the center of the sample to monitor the ambient temperature, denoted as $T_{amb}$. During testing, the sample (3 cm x 3 cm at 0% strain) was stretched to a certain strain via the biaxial actuator, then the testing system was raised to contact the sample by a lab jack. The SEBS side of the sample was facing up. The $T_{obj}$ was set to be 33 °C ($T_{obj} = T_{set} = 33\ °C$) and the $T_1$ was kept the same as $T_2$. All the temperatures were controlled by the DC power supply (RIGOL Technologies, DP831) via the LabVIEW PID program. Additionally, $T_{amb}$ was set to a constant temperature of 25 °C by controlling the circulating water temperature. When all the temperatures are in equilibrium, the interface between the testing heater and PDMS can be regarded as a thermally insulating boundary, and the heat generated from the testing heater can be treated as one-dimensional upward heat flux, denoted as $q$. Subsequently, the heat transfer coefficient ($h$) can be calculated as $h = q/(T_{set} - T_{amb})$. The infrared images in different states were conducted by FLIR E60 compact thermal imaging camera. The top surface temperature was obtained by infrared images and analyzed by ResearchIR software (4.30.2 version, FLIR Systems, Inc.).

**Object temperature measurement and emissivity dependence.**

The measurement of actual temperature change was also conducted using the similar steady-state measurement apparatus of the total heat transfer coefficient measurement. The guard heater was controlled by a LabVIEW PID program to realize the $T_1 = T_2$ (thermally insulating boundary condition). The power of the testing heater was set to a constant power of 100 W/m². Both heaters were powered by the DC power supply (RIGOL Technologies, DP831). $T_{amb}$ was set at 25 °C by controlling the recirculating water temperature. The strain of SEBS/metal film was applied via a lab-made actuator (Actuonix Linear Actuators, P16) insides the testing chamber. The actual temperature of the object ($T_3$) was measured and recorded via a data acquisition (OMB-DAQ-2408 Series, OMEGA) and a LabVIEW program. We have investigated the heat transfer behavior of various heaters with different emissivities via this steady-state measurement apparatus. The emissivity differences were achieved by applying a different coating onto the heaters.

**Construction of pneumatic modulation device.**

A pneumatic thermal camouflage device consists of an inflation/vacuum module, heating module, and infrared appearance module. The gas/vacuum module is used to regulate the air pressure of each pixel. The heat source is provided by the bottom heating plate to mimic the hot object. Infrared appearance module was made of 4 x 4 cylindrical tube pixels (diameter of 5/8" and height of 19/32"), whose top openings were sealed with the SEBS/metal film tightly via cable zip ties. Besides, all these tubes were drilled and mounted with blowholes via epoxy resin. Further, through controlling the air pressure, the *E*-, *R*-, and *T*- mode of SEBS/metal film can be converted and regulated. The infrared images in different statuses were conducted by FLIR E60 compact thermal imaging camera. The pressure-dependent behavior of the film was analyzed in the supplementary information.

## Acknowledgements

The authors thank Pratt School of Engineering at Duke University for the funding support and the Shared Materials Instrumentation Facility (SMIF) for its technical support. H.M.F also acknowledges the China Scholarship Council for their financial support.

## Author contributions

P.-C.H. and H.M.F. conceived the idea. H.M.F. and W.R.X. prepared the triple-mode samples. H.M.F and W.R.X designed and fabricated the biaxial actuator. H.M.F., B.W.S., and Y.T.L. designed and conducted the thermal measurements. W.J.P., H.M.F. and K.B.F conducted the FTIR spectrometry measurement. H.M.F. and X.Q.L tested and simulated mechanical properties. P.-C.H. supervised the project. S.L.B and W.J.P. assisted with the data interpretation. All authors discussed the results and contributed to the writing of the paper.

## Competing interests

P.C.H and H.M.F are listed as inventors on a provisional U.S. patent application from the Duke University, which describes the design and working mechanism of an all-surface radiative thermal managing device. The remaining authors declare no competing interests.